\documentstyle[multicol,aps,epsf]{revtex}
\begin{document}
\draft \title{Lattice Boltzmann Simulations of Liquid Crystal Hydrodynamics}
\author{Colin Denniston$^{1}$, Enzo Orlandini$^{2}$, and J.M. Yeomans$^{1}$}
\address{$^{1}$ Dept. of Physics, Theoretical Physics,
University of Oxford, 1 Keble Road, Oxford OX1 3NP}
\address{$^{2}$ INFM-Dipartimento di Fisica, Universit\`a di
  Padova, 1-35131 Padova, Italy}
\date{\today} \maketitle

\begin{abstract}
We describe a lattice Boltzmann algorithm to simulate liquid crystal
hydrodynamics. The equations of motion are written in terms of a
tensor order parameter. This allows both the isotropic and the nematic
phases to be considered. Backflow effects and the hydrodynamics of
topological defects are naturally included in the simulations, as are
viscoelastic properties such as shear-thinning and shear-banding.
\end{abstract}

\renewcommand{\theequation}{\thesection.\arabic{equation}}
\renewcommand{\baselinestretch}{1.2} \def\xbar{1/x} \def\ra{\rangle}
\def\la{\langle} \def\Pr{{\it Proof: }} \def\qed{$\Box$} \def\cF{{\cal
F}} \def\cFo{{\cal F}^o} \def\pan{\par\noindent}
\def\pa{\partial_{\alpha}} \def\pb{\partial_{\beta}}
\def\pe{\partial_{\eta}} \def\pt{\partial_t
} \def\eia{e_{i\alpha}}
\def\eib{e_{i\beta}} \def\eie{e_{i\eta}} \def\geq{g_i^{eq}}
\def\feq{f_i^{eq}} \def\dt{\Delta t} \def\ua{u_{\alpha}}
\def\ue{u_{\eta}} \def\Tea{T_{\eta\alpha}} \def\Pea{P_{\eta\alpha}}

\section{Introduction}

Liquid crystalline materials are often made up of long, thin, rod-like 
molecules\cite{DG93}.  The molecular geometry and interactions can lead to 
a wide range of equilibrium phases.  Here we shall be concerned 
with two of the simplest, the isotropic phase, where the 
orientation of the molecules is random, and the nematic phase, 
where the molecules tend to align along a preferred direction.

The aim of this paper is to describe a numerical scheme which 
can explore the hydrodynamics of liquid crystals within both 
the isotropic and the nematic phases.  There are two major 
differences between the hydrodynamics of simple liquids and 
that of liquid crystals.  First, the geometry of the 
molecules means that they are rotated by gradients in the 
velocity field.  Second, the equilibrium free energy is more complex 
than for a simple fluid and this in turn increases the 
complexity of the stress tensor in the
Navier-Stokes equation for the evolution of the fluid 
momentum.  This coupling between the elastic energy and 
the flow leads to rich hydrodynamic behaviour.  A simple 
example is the existence of a tumbling phase where the molecules 
rotate in an applied shear\cite{RT98}. Other examples include
shear banding, a non-equilibrium phase separation 
into coexisting states with different strain rates\cite{MR97}, and the
possibility of Williams domains, convection 
cells induced by an applied electric field\cite{DG93}.

The equations of motion describing liquid crystal 
hydrodynamics are complex.  There are several derivations
broadly in agreement, but differing in the detailed form 
of some terms.  Here we follow the approach of Beris and 
Edwards\cite{BE94} who write the equations of 
motion in terms of a tensor order parameter ${\bf Q}$ 
which can be related to the second moment of the orientational
distribution function of  
the molecules.  This has the advantage that the hydrodynamics of 
both the isotropic and the nematic phases, and of topological defects
in the nematic phase, can be included within the same formalism.  
Most other theories of liquid crystal hydrodynamics appear 
as limiting cases.  In particular the Ericksen-Leslie formulation of
nematodynamics\cite{E66,L68}, widely used in 
the experimental liquid crystal literature, follows when 
uniaxiality is imposed and the magnitude of the order parameter is
held constant. 

Considerable analytic progress in understanding liquid crystal 
flow in simple geometries has been made, but this is 
inevitably limited by the complexity of the equations of 
motion.  Therefore it is useful to formulate a method of 
obtaining numerical solutions of the
 hydrodynamic equations to further explore their rich 
phenomenology.  Moreover we should like to be able 
to predict flow patterns for given 
viscous and elastic coefficients for comparison to experiments and 
to explore the effects of hydrodynamics when liquid crystals are 
used in display devices or during industrial processing.  

Rey and Tsuji\cite{RT98} have obtained interesting results on flow-induced
ordering of the director field and on defect dynamics by solving the
Beris-Edwards equation for the order parameter. However, the velocity
field was imposed externally and no back-flows (effect of the
director configuration on the velocity field) were included. Fukuda\cite{F98}
used an Euler scheme to solve a model somewhat simpler than the full
Beris-Edwards model but still including backflow, 
and studied the effect of hydrodynamics on phase ordering
in liquid crystals. Otherwise
most previous work on liquid crystal hydrodynamics has been limited
to a constant order parameter (the Ericksen-Leslie-Parodi equations)
and often restricted to one dimension. 

Lattice Boltzmann schemes have recently proved very 
successful in simulations of complex fluids and it is 
this approach that we shall take here\cite{CD98}. Such algorithms
can be usefully and variously considered as a slightly 
unusual finite-difference discretization of the equations 
of motion or as a lattice 
version of a simplified Boltzmann equation.
It is not understood why 
the approach is particularly useful for complex fluids but it may 
be related to the very natural way in which a free energy 
describing the equilibrium properties of the fluid can be 
incorporated in the simulations, drawing on ideas from 
statistical mechanics\cite{SO96}.  
Recent applications have included phase ordering and flow 
in binary fluids\cite{Y99} and self-assembly and spontaneous emulsification in 
amphiphilic fluids\cite{LG99,TG99}.

However, in applications so far, with the exception of \cite{care}, 
the order parameter has been a scalar and has coupled to 
the flow via a simple advective term.  The liquid crystal equations of 
motion are written in terms of a tensor order parameter. 
This is responsible for the main new features of 
the lattice Boltzmann approach described in this paper. It also leads
to the possibility of exploring viscoelastic fluid behaviour such as
shear-thinning and shear-banding without the
need to impose a constitutive equation for the stress\cite{L88}.

In Section 2 we summarise the hydrodynamic equations of 
motion for liquid crystals.
The lattice Boltzmann scheme is defined in Section 3. A modified
version of the collision operator is used to eliminate lattice
viscosity effects.   
Section 4 describes a Chapman-Enskog expansion which relates the 
numerical scheme to the hydrodynamic equations of motion.  
Numerical results for simple shear flows are presented in 
Section 5 and other possible applications of the approach are 
outlined in Section 6.

\section{The hydrodynamic equations of motion}
\label{2.0}

We shall follow the formulation of liquid crystal hydrodynamics
described by Beris and Edwards\cite{BE94}. The continuum 
equations of motion are written
in terms of a tensor order parameter ${\bf Q}$ which is related to the
direction of individual molecules $\vec {\hat{n}}$ by 
$Q_{\alpha\beta}= \langle \hat{n}_\alpha \hat{n}_\beta -
{1\over 3} \delta_{\alpha\beta}\rangle$ where the angular brackets
denote a coarse-grained average. (Greek indices will be
used to represent Cartesian components of vectors and tensors 
and the usual summation over
repeated indices will be assumed.) ${\bf Q}$ is a traceless symmetric
tensor which is zero in the isotropic
phase. We first write down a Landau
free energy which describes the equilibrium properties of the liquid 
crystal and the isotropic--nematic transition. This appears in the 
equation of motion of the order parameter, which includes a
Cahn-Hilliard-like term through which the system evolves towards 
thermodynamic equilibrium. It also includes a
term coupling the order parameter to the flow. The order parameter is
both advected by the flow and, because liquid crystal molecules
are rod-like, rotated by velocity gradients.

We then write down the continuity and Navier-Stokes equations for the
evolution of the flow field. In particular the form of the stress
appropriate to a tensor order parameter is discussed. A brief
comparison is given to a similar formalism introduced by Doi\cite{D81}
and extended by Olmsted {\it et.\ al.}\cite{OG92,OD97}.
For a uniaxial nematic in the absence of 
any defects the Beris-Edwards equations reduce to the Ericksen-Leslie-Parodi
formulation of nematodynamics\cite{DG93}. The hydrodynamic behaviour
of nematic liquid crystals is often characterised in terms of the
Leslie coefficients and it is therefore useful to list them below. More
details of the
mapping between the Beris-Edwards and the Ericksen-Leslie-Parodi
equations are given in Appendix A.\\
{\bf Free energy:}
The equilibrium properties of a liquid crystal in solution can be
described by a free energy\cite{OD97}
\begin{equation}
{\cal F}=\int d^3 r \left\{ \frac{a}{2} 
  Q_{\alpha \beta}^2 - \frac{b}{3} Q_{\alpha \beta}
Q_{\beta \gamma}Q_{\gamma \alpha}+ \frac{c}{4}
  (Q_{\alpha \beta}^2)^2 + \frac{\kappa}{2} (\partial_\alpha Q_{\beta \lambda})^2
  \right\}.
\label{free}
\end{equation}
We shall work within the one elastic constant
approximation. Although it is not hard to include more general
elastic terms this simplification will not affect the qualitative
behaviour. The free energy (\ref{free}) describes a first order transition from the isotropic to the nematic
phase.\\
{\bf Equation of motion of the nematic order parameter:}
The equation of motion for the nematic order parameter is\cite{BE94}
\begin{equation}
(\partial_t+{\vec u}\cdot{\bf \nabla}){\bf Q}-{\bf S}({\bf W},{\bf
  Q})= \Gamma {\bf H}
\label{Qevolution}
\end{equation}
where $\Gamma$ is a collective rotational diffusion constant.
The first term on the left-hand side of equation (\ref{Qevolution})
is the material derivative describing the usual time dependence of a
quantity advected by a fluid with velocity ${\vec u}$. This is
generalised by a second term 
\begin{eqnarray}
{\bf S}({\bf W},{\bf Q})
&=&(\xi{\bf D}+{\bf \Omega})({\bf Q}+{\bf I}/3)+({\bf Q}+
{\bf I}/3)(\xi{\bf D}-{\bf \Omega})\nonumber\\
& & -2\xi({\bf Q}+{\bf I}/3){\mbox{Tr}}({\bf Q}{\bf W})
\end{eqnarray}
where ${\bf D}=({\bf W}+{\bf W}^T)/2$ and
${\bf \Omega}=({\bf W}-{\bf W}^T)/2$
are the symmetric part and the anti-symmetric part respectively of the
velocity gradient tensor $W_{\alpha\beta}=\partial_\beta u_\alpha$.
${\bf S}({\bf W},{\bf Q})$  appears in the equation of motion because
the order parameter distribution can be both rotated and stretched by
flow gradients. $\xi$ is a constant which will depend on the molecular
details of a given liquid crystal.

The term on the right-hand side of equation (\ref{Qevolution})
describes the relaxation of the order parameter towards the minimum of
the free energy. The molecular field ${\bf H}$ which provides the driving
motion is related to the derivative of the free energy by
\begin{eqnarray}
{\bf H}&=& -{\delta {\cal F} \over \delta Q}+({\bf
    I}/3) Tr{\delta {\cal F} \over \delta Q}\nonumber\\ &=&
    -a {\bf Q}+ b \left({\bf Q^2}-({\bf
    I}/3)Tr{\bf Q^2}\right)- c  {\bf Q}Tr{\bf Q^2}+\kappa \nabla^2
    {\bf Q}.
\label{H(Q)}
\end{eqnarray}  \\
{\bf Continuity and Navier-Stokes equations:}
The fluid momentum obeys the continuity
\begin{equation}
\pt \rho + \pa \rho u_{\alpha} =0,
\label{continuity}
\end{equation}
where $\rho$ is the fluid density, and the Navier-Stokes equation 
\begin{equation}
 \rho\partial_t u_\alpha+\rho u_\beta \partial_\beta
u_\alpha=\partial_\beta \tau_{\alpha\beta}+\partial_\beta
\sigma_{\alpha\beta}+{\rho \tau_f \over
3}(\partial_\beta((\delta_{\alpha \beta}-3\partial_\rho
P_{0})\partial_\gamma u_\gamma+\partial_\alpha
u_\beta+\partial_\beta u_\alpha).
\label{NS}
\end{equation}
The form of the equation is not dissimilar to that for a simple
fluid. However the details of the stress tensor reflect the additional
complications of liquid crystal hydrodynamics. 
There is a symmetric contribution
\begin{eqnarray}
\sigma_{\alpha\beta} &=&-P_0 \delta_{\alpha \beta}
-\xi H_{\alpha\gamma}(Q_{\gamma\beta}+{1\over
  3}\delta_{\gamma\beta})-\xi (Q_{\alpha\gamma}+{1\over
  3}\delta_{\alpha\gamma})H_{\gamma\beta}\nonumber\\
& & \quad +2\xi
(Q_{\alpha\beta}+{1\over 3}\delta_{\alpha\beta})Q_{\gamma\epsilon}
H_{\gamma\epsilon}-\partial_\beta Q_{\gamma\nu} {\delta
{\cal F}\over \delta\partial_\alpha Q_{\gamma\nu}}
\label{BEstress}
\end{eqnarray}
and an antisymmetric contribution
\begin{equation}
 \tau_{\alpha \beta} = Q_{\alpha \gamma} H_{\gamma \beta} -H_{\alpha
 \gamma}Q_{\gamma \beta} .
\label{as}
\end{equation}
The pressure $P_0$ is taken to be
\begin{equation}
P_0 = \rho T -{\kappa \over 2}(\nabla{\bf Q})^2.
\end{equation}

An earlier development of liquid crystal hydrodynamics in terms of a
tensor order parameter was proposed by Doi\cite{D81}. The Doi theory is based
upon a Smoluchowski evolution equation (similar to the Boltzmann
equation for translational motion) for the orientational distribution
function. The main advantage of the approach is the possibility of
relating the phenomenological coefficients in the equations of motion
to microscopic parameters. One omission is the lack of gradient terms
in the free energy (but see \cite{OD97}). Moreover it is necessary to
use closure
approximations to obtain a tractable set of hydrodynamic
equations. The Doi and Beris--Edwards equations are very similar: the
main difference is in the symmetric contribution to the stress
tensor. The Doi theory gives a simpler form which is incomplete in
that it does not obey Onsager reciprocity. (A similar comment applies
to all closure relations that we have found in the literature.)

Hydrodynamic equations for the nematic phase were formulated by
Ericksen and Leslie\cite{E66,L68,DG93}. These are widely used as the Leslie coefficients
provide a useful measure of the viscous properties of the liquid
crystal fluid. The Beris-Edwards equations reduce to those of Ericksen
and Leslie in the uniaxial nematic phase when the magnitude of the
order parameter remains constant. Hence a limitation of the
Ericksen-Leslie theory is
that it cannot include the hydrodynamics of topological defects. For
convenience we list below the relationship between the Leslie
coefficients and the parameters appearing in the equations of motion
(\ref{Qevolution}) and (\ref{NS}). An outline of their derivation from
the Beris--Edwards approach is given in Appendix A.
\begin{eqnarray} 
\alpha_1&=& -\frac{2}{3} q^2(3+4 q-4 q^2)\xi^2 / \Gamma \label{EL1}\\
  \alpha_2&=& (-\frac{1}{3}q(2+q)\xi-q^2) / \Gamma \\
  \alpha_3&=& (-\frac{1}{3}q(2+q)\xi+q^2) / \Gamma \\
  \alpha_4&=& \frac{4}{9}(1-q)^2\xi^2 / \Gamma + \eta \\
  \alpha_5&=& (\frac{1}{3}q(4-q)\xi^2+\frac{1}{3}q(2+q)\xi) / \Gamma \\
  \alpha_6&=& (\frac{1}{3}q(4-q)\xi^2-\frac{1}{3}q(2+q)\xi) / \Gamma
\label{EL6} 
  \end{eqnarray} 
where $q$ is the magnitude of the nematic order parameter and
$\eta=\rho \tau_f/3$.

A detailed comparison of the theories of liquid crystal hydrodynamics
can be found in Beris and Edwards\cite{BE94}.

\section{A lattice Boltzmann algorithm for liquid crystal
hydrodynamics}

We now define a lattice Boltzmann algorithm which solves the
hydrodynamic equations of motion of a liquid crystal 
(\ref{Qevolution}), (\ref{continuity}), and (\ref{NS}). 
Lattice Boltzmann algorithms are defined in
terms of a set of continuous variables, usefully termed partial
distribution functions, which move on a lattice in discrete space and
time. They were first developed as mean-field versions of cellular
automata simulations but can also usefully be viewed as a particular
finite-difference implementation of the continuum equations of motion\cite{CD98}.

Lattice Boltzmann approaches have been particularly successful in
modeling fluids which evolve to minimise a free energy\cite{SO96}. It is not
proven why this is the case, but one can surmise that the existence of
an H-theorem, which governs the approach to equilibrium, helps to
enhance the stability of the scheme\cite{W98,KF99}.

The simplest lattice Boltzmann algorithm, which describes the
Navier-Stokes equations of a simple fluid, is defined in terms of a
single set of partial distribution functions which sum on each site to
give the density. For liquid crystal hydrodynamics this must be
supplemented by a second set, which are tensor variables, and which
are related to the tensor order parameter ${\bf Q}$. A
description of the algorithm is given in Section \ref{C2} and the
continuum limit is taken in Section \ref{C3}. A
Chapman-Enskog expansion\cite{CC90} showing how the algorithm
reproduces the liquid crystal equations of motion follows in Section \ref{C4}.

\subsection{The lattice Boltzmann algorithm}
\label{C2}

We define two distribution functions, the scalars $f_i (\vec{x})$ and
the symmetric traceless tensors ${\bf G}_i (\vec{x})$ on each lattice
site $\vec{x}$. Each $f_i$, ${\bf G}_i$ is associated with a lattice
vector ${\vec e}_i$. We choose a nine-velocity model on a square
lattice with velocity vectors ${\vec e}_i=(\pm 1,0),(0,\pm 1), (\pm 1, \pm
1), (0,0)$. Physical variables are defined as moments of the
distribution function
\begin{equation}
\rho=\sum_i f_i, \qquad \rho u_\alpha = \sum_i f_i  e_{i\alpha},
\qquad {\bf Q} = \sum_i {\bf G}_i.
\label{eq1}
\end{equation} 

The distribution functions evolve in a time step $\Delta t$ according
to
\begin{eqnarray}
&&f_i({\vec x}+{\vec e}_i \Delta t,t+\Delta t)-f_i({\vec x},t)=
\frac{\Delta t}{2} \left[{\cal C}_{fi}({\vec x},t,\left\{f_i
\right\})+ {\cal C}_{fi}({\vec x}+{\vec e}_i \Delta
t,t+\Delta
t,\left\{f_i^*\right\})\right],
\label{eq2}\\
&&{\bf G}_i({\vec x}+{\vec e}_i \Delta t,t+\Delta t)-{\bf G}_i({\vec
x},t)= \nonumber\\
&& \qquad\qquad \qquad\qquad \qquad\qquad\frac{\Delta t}{2}\left[ {\cal C}_{{\bf G}i}({\vec
x},t,\left\{{\bf G}_i \right\})+
                {\cal C}_{{\bf G}i}({\vec x}+{\vec e}_i \Delta
                t,t+\Delta t,\left\{{\bf G}_i^*\right\})\right].
\label{eq3}
\end{eqnarray}
This represents free streaming with velocity ${\vec e}_i$ and a
collision step which allows the distribution to relax towards
equilibrium. 
$f_i^*$ and ${\bf G}_i^*$ are first order approximations to 
$f_i({\vec x}+{\vec e}_i \dt,t+\dt)$ and ${\bf G}_i({\vec x}+{\vec e}_i \dt,t+\dt)$
respectively. They are obtained from equations
(\ref{eq2}) and (\ref{eq3}) but with $f_i^*$ and ${\bf G}_i^*$ set to
$f_i$ and ${\bf G}_i$.
Discretizing in this way, which is similar to a predictor-corrector 
scheme, has the advantages that lattice viscosity terms are eliminated
to second order and that the stability of the scheme is improved.

The collision operators are taken to have the form of a single
relaxation time Boltzmann equation\cite{CD98}, together with a forcing term
\begin{eqnarray}
{\cal C}_{fi}({\vec x},t,\left\{f_i \right\})&=&
-\frac{1}{\tau_f}(f_i({\vec x},t)-f_i^{eq}({\vec x},t,\left\{f_i
\right\}))
+p_i({\vec x},t,\left\{f_i \right\}),
\label{eq4}\\
{\cal C}_{{\bf G}i}({\vec x},t,\left\{{\bf G}_i
\right\})&=&-\frac{1}{\tau_{g}}({\bf G}_i({\vec x},t)-{\bf
G}_i^{eq}({\vec x},t,\left\{{\bf G}_i \right\}))
+{\bf M}_i({\vec x},t,\left\{{\bf G}_i \right\}).
\label{eq5}
\end{eqnarray}

The form of the equations of motion and thermodynamic equilibrium
follow from the choice of the moments of the equilibrium distributions
$f^{eq}_i$ and ${\bf G}^{eq}_i$ and the driving terms $p_i$ and
${\bf M}_i$. $f_i^{eq}$ is constrained by
\begin{equation}
\sum_i f_i^{eq} = \rho,\qquad \sum_i f_i^{eq} e_{i \alpha} = \rho
u_{\alpha}, \qquad
\sum_i f_i^{eq} e_{i\alpha}e_{i\beta} = -\sigma_{\alpha\beta}+\rho
u_\alpha u_\beta
\label{eq6} 
\end{equation}
where the zeroth and first moments are chosen to impose conservation
of
mass and momentum. The second moment of $f^{eq}$ controls the symmetric
part of the stress tensor, whereas the moments of $p_i$
\begin{equation}
\sum_i p_i = 0, \quad \sum_i p_i e_{i\alpha} = \partial_\beta
\tau_{\alpha\beta},\quad \sum_i p_i
e_{i\alpha}e_{i\beta} = 0
\label{eq7}
\end{equation}
impose the antisymmetric part of the stress tensor.
For the equilibrium of the order parameter distribution we choose
\begin{equation}
\sum_i {\bf G}_i^{eq} = {\bf Q},\qquad \sum_i
{\bf G}_i^{eq} {e_{i\alpha}} = {\bf Q}{u_{\alpha}},
\qquad \sum_i {\bf G}_i^{eq}
e_{i\alpha}e_{i\beta} = {\bf Q} u_\alpha u_\beta .
\label{eq8}
\end{equation}
This ensures that the order parameter
is convected with the flow. Finally the evolution of the
order parameter is most conveniently modeled by choosing
\begin{equation}
\sum_i {\bf M}_i = \Gamma {\bf H}({\bf Q})
+{\bf S}({\bf W},{\bf Q}) \equiv {\bf \hat{H}}, \qquad
\qquad \sum_i {\bf M}_i {e_{i\alpha}} = (\sum_i {\bf M}_i)
{u_{\alpha}}.
\label{eq9}
\end{equation}
which ensures that the fluid minimises its free energy at equilibrium.

Conditions (\ref{eq6})--(\ref{eq9})
can be satisfied as is usual in lattice Boltzmann
schemes by writing the equilibrium distribution functions and forcing
terms as polynomial expansions in the velocity\cite{CD98}
\begin{eqnarray}
f_i^{eq}&=&A_s + B_s u_\alpha e_{i\alpha}+C_s u^2+D_s u_\alpha
u_\beta
e_{i\alpha}e_{i\beta}+E_{s\alpha\beta}e_{i\alpha}e_{i\beta},\nonumber \\
{\bf G}_i^{eq}&=&{\bf J}_s + {\bf K}_s u_\alpha e_{i\alpha}+{\bf L}_s
u^2+{\bf N}_s u_\alpha
u_\beta e_{i\alpha}e_{i\beta},\nonumber \\ 
p_i&=&T_s \partial_\beta \tau_{\alpha\beta} e_{i\alpha},\nonumber \\ 
{\bf M}_i&=&{\bf R}_s+{\bf S}_s u_\alpha e_{i\alpha},
\end{eqnarray}
where $s={\vec e}_i\;^2 \in \{0,1,2\}$ identifies separate coefficients
for different absolute values of the velocities.  A suitable choice is
\begin{eqnarray}
&&A_2=(\sigma_{xx}+\sigma_{yy})/16,\qquad A_1=2 A_2,\qquad A_0=\rho-12 A_2,\nonumber \\
&&B_2=\rho/12,\qquad B_1=4 B_2,\nonumber \\ &&C_2=-\rho/16,\qquad C_1=-\rho/8,
\qquad C_0=-3 \rho/4,\nonumber \\ &&D_2=\rho/8, \qquad D_1=\rho/2\nonumber \\
&&E_{2xx}=(\sigma_{xx}-\sigma_{yy})/16,\qquad E_{2yy}=-E_{2xx}, \qquad
E_{2xy}=E_{2yx}=\sigma_{xy}/8,\nonumber \\ &&E_{1xx}=4 E_{2xx},\qquad E_{1yy}=4
E_{2yy},\nonumber \\ &&{\bf J}_0={\bf Q},\nonumber \\ 
&&{\bf K}_2={\bf Q}/12,\qquad {\bf K}_1=4 {\bf K}_2,\nonumber \\
&&{\bf L}_2=-{\bf Q}/16,\qquad {\bf L}_1=-{\bf Q}/8, \qquad {\bf
L}_0=-3 {\bf Q}/4,\nonumber \\ 
&&{\bf N}_2={\bf Q}/8, \qquad
{\bf N}_1={\bf Q}/2\nonumber \\ 
&&T_2=1/12, \qquad T_1=4 T_2,\nonumber \\ 
&&{\bf R}_2=\widehat{\bf H}/9, \qquad
{\bf R}_1={\bf R}_0={\bf R}_2\nonumber \\ &&{\bf S}_2=\widehat{\bf H}/12, \qquad {\bf S}_1=4 {\bf S}_2,
\label{eq14}
\end{eqnarray}
where any coefficients not listed are zero.

\subsection{Continuum limit}
\label{C3}

We write down the continuum limit of the lattice Boltzmann evolution
equations (\ref{eq2}) and (\ref{eq3}) showing, in particular, that the
predictor-corrector form of the collision integral eliminates lattice
viscosity effects to second order.

Consider equation (\ref{eq2}). 
Taylor expanding $f_i({\vec x}+{\vec e}_i \dt,t+\dt)$ gives
\begin{equation}
f_i({\vec x}+{\vec e}_i \dt,t+\dt) = f_i({\vec x},t)+ \dt D f_i({\vec
  x},t)+{\dt^2 \over 2} D^2 f_i({\vec x},t) +O(\dt^3)
\end{equation}
where $D \equiv \pt +e_{i\alpha} \pa$. Similarly,
expanding the collision term equation(\ref{eq4}),
\begin{eqnarray}
& & {\cal C}_{fi}({\vec x}+{\vec e}_i \dt,t+\dt,\left\{f_i +\dt {\cal
  C}_{fi}({\vec x},t,\left\{f_i \right\})\right\})={\cal C}_{fi}({\vec
  x},t,\left\{f_i \right\})+\nonumber\\ & & \qquad\qquad \dt D {\cal
  C}_{fi}({\vec x},t,\left\{f_i \right\})
+ O(\dt^2)
\end{eqnarray}
and substituting into equation (\ref{eq2}) gives
\begin{equation}
D f_i({\vec x},t) = {\cal C}_{fi}({\vec x},t,\left\{f_i \right\})
 -{\dt \over 2} \left\{ D^2 f_i({\vec x},t) 
-D {\cal C}_{fi}({\vec x},t,\left\{f_i \right\})
\right\} + O(\dt^2).
\label{eq17}
\end{equation}
We see immediately that
\begin{equation}
D f_i({\vec x},t) = {\cal C}_{fi}({\vec x},t,\left\{f_i \right\})+O(\dt).
\label{eq18}
\end{equation}
Using equation(\ref{eq18}) in the expansion (\ref{eq17}) it follows
that there are no terms of order $\dt$ in (\ref{eq17}) and
\begin{equation}
D f_i({\vec x},t) = {\cal C}_{fi}({\vec x},t,\left\{f_i \right\})+O(\dt^2).
\label{eq19}
\end{equation}
A similar expansion of equation (\ref{eq3}) leads to
\begin{equation}
D {\bf G}_i({\vec x},t) = {\cal C}_{{\bf G}i}({\vec x},t,\left\{{\bf G}_i
\right\})+O(\dt^2).
\label{eq20}
\end{equation}

In the standard lattice Boltzmann discretization terms of order $\dt$
appear in equations (\ref{eq19}) and (\ref{eq20}). These are of
similar forms to those which arise from the Chapman-Enskog expansion
and have been subsumed into the viscosity. However this is 
not generally possible and
it is convenient to use the predictor-corrector form for the collision
term assumed in equations (\ref{eq4}) and (\ref{eq5}) to eliminate
them at this stage.

\subsection{Chapman-Enskog expansion}
\label{C4}

We can now proceed with a Chapman-Enskog expansion, an expansion of
the distribution functions about equilibrium, which assumes that
successive derivatives are of increasingly high order\cite{CC90}. The aim is to
show that equation (\ref{eq20}) reproduces the evolution equation of
the liquid crystal order parameter (\ref{Qevolution}) and equation
(\ref{eq19}) the continuity and Navier-Stokes equations
(\ref{continuity}) and (\ref{NS}) to second order in derivatives.
Writing
\begin{equation}
{\bf G}_i= {\bf G}_i^{(0)}+ {\bf G}_i^{(1)}+ {\bf G}_i^{(2)}+\ldots
\label{eq21}
\end{equation}
and substituting into (\ref{eq20}) using the form for the collision term
(\ref{eq5}) gives, to zeroth order
\begin{equation}
{\bf G}_i^{(0)}={\bf G}_i^{eq}+\tau_g {\bf M}_i.
\end{equation}
Summing over $i$ and using, from equations (\ref{eq1}) and (\ref{eq8}),
\begin{equation}
\sum_i {\bf G}_i \equiv {\bf Q} = \sum_i {\bf G}_i^{eq}
\label{eq23}
\end{equation}
shows that the zeroth moment of ${\bf M}_i$ appears at first order in the
Chapman-Enskog expansion. This is as expected because, from equation
(\ref{eq9}), $\sum_i {\bf M}_i$ is
related to free energy derivatives which will be zero in
equilibrium.  The first moment will 
also be first order in derivatives.

It then follows, from substituting equation (\ref{eq21}) into equation
(\ref{eq20}), that the first and second order deviations of the
distribution function from equilibrium are
\begin{equation}
{\bf G}_i^{(1)} =  - \tau_g D {\bf G}_i^{eq} + \tau_g {\bf M}_i,
\label{eq24}
\end{equation}
\begin{equation}
{\bf G}_i^{(2)} = \tau_g^2 D^2 {\bf G}_i^{eq} - \tau_g^2 D {\bf M}_i.
\label{eq25}
\end{equation}
Using equation (\ref{eq24}) in equation (\ref{eq21}), summing over $i$
and using (\ref{eq23}), (\ref{eq8}), and (\ref{eq9}) gives, to
first order,
\begin{equation}
\pt {\bf Q} + \pa ({\bf Q} \ua ) = \widehat{\bf H} + O(\partial^2)
\label{eq26}
\end{equation}

The second order term (\ref{eq25}) gives, after a lengthy calculation,
described in Appendix B, a correction
\begin{equation}
-\tau_g \left(\pa\left(\frac{\bf Q}{\rho}\pe P_0 \right) \right ).
\label{eq27}
\end{equation}
This additional term is a feature common to most lattice Boltzmann
models of complex fluids. It is not known whether it has a physical
orign, but it is very small in all the cases tested so far and has no
effect upon the behaviour of the fluid.

A similar expansion for the partial density distribution functions
$f_i$ gives the continuity and Navier-Stokes equations. Writing
\begin{equation}
f_i=f_i^{(0)}+f_i^{(1)}+f_i^{(2)}+\ldots,
\label{eq28}
\end{equation}
substituting into (\ref{eq19}) and using the collision operator (\ref{eq4}) gives
\begin{equation}
f_i^{(0)}=\feq+\tau_f p_i,
\label{eq29}
\end{equation}
\begin{equation}
f_i^{(1)}=  - \tau_f D \feq - \tau_f^2 D p_i,
\label{eq30}
\end{equation}
\begin{equation}
f_i^{(2)} = \tau_f^2 D^2\feq +  \tau_f^3 D^2 p_i.
\label{eq31}
\end{equation}
Summing $f_i$ over $i$ and using the constraints on the moments
of $f_i$, $\feq$ and $p_i$, from equations (\ref{eq1}), (\ref{eq6}) and (\ref{eq7})
respectively
\begin{eqnarray}
\left ( \pt \rho + \pa \rho \ua +\tau_f \pa\sum_i
  p_i\eia\right)&=&\tau_f \pt \left[\pt \rho + \pa \rho \ua
  +\tau_f \pa\sum_i p_i\eia\right]\nonumber\\ &+& \tau_f \pa \left[
  \pt \rho \ua + \pb\sum_i\feq\eia\eib + \tau_f \pt\sum_i p_i\eia
  \right].
\label{eq32}
\end{eqnarray}
The first term in square brackets is second order in derivatives. Therefore
\begin{equation}
\left ( \pt \rho + \pa \rho \ua +\tau_f \pa\sum_i p_i\eia\right)=
  \tau_f \pa \left[ \pt \rho \ua + \pb\sum_i\feq\eia\eib + \tau_f
  \pt\sum_i p_i\eia \right]+O(\partial^3).
\label{eq33}
\end{equation}

We now multiply Eq.(\ref{eq28}) by $e_{i\alpha}$ and sum over $i$.
Using the constraints (\ref{eq6}) and (\ref{eq7}) 
and the definitions (\ref{eq1}) 
\begin{eqnarray}
&&\left ( \pt \rho\ua + \pb \sum_i\feq\eia\eib +\tau_f \pt\sum_i
  p_i\eia\right) = \nonumber \\
&&\qquad\qquad \sum_i p_i\eia+\tau_f \pt \left[\pt \rho\ua +
  \pb \sum_i\feq\eia\eib +\tau_f \pt\sum_i p_i\eia \right]\nonumber\\
  &&\qquad\qquad + \tau_f \pb \left[ \pt \sum_i\feq\eia\eib +
  \partial_\gamma \sum_i\feq\eia\eib e_{i\gamma} + \tau_f
  \partial_\gamma \sum_i p_i\eia\eib e_{i\gamma} \right].
\label{eq34}
\end{eqnarray}
So to first order in derivatives 
\begin{equation}
\left ( \pt \rho\ua + \pb \sum_i\feq\eia\eib +\tau_f \pt\sum_i
  p_i\eia\right) = \sum_i p_i\eia+O(\partial^2).
\label{eq35}
\end{equation}
Placing (\ref{eq35}) into the square brackets in equation (\ref{eq33})
we obtain the continuity equation (\ref{continuity}) to second order in derivatives
\begin{equation}
\left ( \pt \rho + \pa \rho \ua \right)= 0+O(\partial^3).
\label{eq36}
\end{equation}

Substituting equation (\ref{eq35}) into the first square brackets in
equation (\ref{eq34}) and imposing the constraints on the first
moment of the $p_i$ and the second moment of the $f_i^{eq}$, equations
(\ref{eq7}) and (\ref{eq6}), gives
\begin{eqnarray}
& &\pt (\rho\ua) + \pb( \rho \ua u_\beta) = \pb
 \sigma_{\alpha\beta}+\pb \tau_{\alpha\beta} \nonumber \\ 
& & \qquad\qquad +\tau_f \pb \left[ -\pt
 \sigma_{\alpha\beta}+\pt (\rho \ua u_\beta) + \partial_\gamma
 \sum_i\feq\eia\eib e_{i\gamma} + \tau_f \partial_\gamma \sum_i
 p_i\eia\eib e_{i\gamma} \right]
\label{eq37}
\end{eqnarray}
showing immediately that the equation of motion (\ref{NS}) is reproduced
to Euler level (first order in derivatives).

From the definitions (\ref{eq14})
\begin{eqnarray}
\sum_i \feq\eia\eib e_{i\gamma}&=&{\rho \over 3}(u_\alpha
\delta_{\beta\gamma}+u_\beta \delta_{\alpha\gamma}+u_\gamma
\delta_{\alpha\beta}),  \label{eq90}\\ 
\sum_i p_i\eia\eib e_{i\gamma}&=&{1 \over
3}(\partial_\delta \tau_{\delta\alpha} \delta_{\beta\gamma}
+\partial_\delta \tau_{\delta\beta} \delta_{\alpha\gamma}
+\partial_\delta \tau_{\delta\gamma} \delta_{\alpha\beta}).
\label{eq91}
\end{eqnarray}
Using equations (\ref{eq90}) and (\ref{eq91}) the viscous
terms in the square brackets in equation (\ref{eq37}) can be
simplified. We assume that the fluid is incompressible, ignore terms
of third order in the velocities, and furthermore assume that, within
these second order terms, the stress tensor can be approximated by
minus the
equilibrium pressure $P_0$. We consider each term in the square
brackets in turn:
\begin{enumerate}
\item
The first term can be rewritten as
\begin{equation}
\partial_t \sigma_{\alpha\beta}=-(\partial_\rho P_{0})(\pt
\rho)\delta_{\alpha\beta}=\rho(\partial_\rho P_{0}) \partial_{\gamma}
u_{\gamma}\delta_{\alpha \beta}
\end{equation}
where the last step follows using the continuity equation (\ref{continuity}). 
\item
Rewriting
\begin{equation}
\partial_t (\rho u_{\alpha} u_{\beta})= \partial_t (\rho u_{\alpha})
u_{\beta} +u_{\alpha} \partial_t (\rho u_{\beta})
\end{equation}
and replacing time derivatives with space derivatives using the Euler
terms in equation (\ref{eq37}) one sees that this term is zero, given
the assumptions listed above.
\item
Using equation (\ref{eq90})
\begin{equation}
\partial_{\gamma} \sum_i \feq\eia\eib e_{i\gamma}={\rho \over
3}(\partial_{\beta} u_\alpha
+\partial_{\alpha} u_\beta + \partial_{\gamma} u_\gamma
\delta_{\alpha\beta})
\end {equation} 
\item
From equation (\ref{eq91}) the fourth
term is of third order in derivatives and can be neglected.
\end{enumerate}
Replacing the square brackets in the equation (\ref{eq37}) with 
the contributions from $1$ and $3$ we obtain the incompressible
Navier-Stokes equation (\ref{NS}).

\section{Numerical Results}

The primary aim of this paper is to describe the details of a
numerical algorithm for simulating liquid crystal hydrodynamics.  
Therefore we restrict ourselves here to presenting a few, brief, 
test cases, aimed at checking the approach.  Further 
numerical applications are listed in the summary of 
the paper and will be presented in detail elsewhere.

In equilibrium with no flow the free energy (\ref{free}) is minimised.
For a generic lyotropic liquid crystal we take
$a=(1-\gamma/3)$ and $b=c=\gamma$, where $\gamma=\phi L\nu_2/\alpha$ is
Doi's excluded volume parameter 
\cite{D81,BE94}.
($L$ is the molecular aspect ratio, $\phi$ the concentration, and
$\nu_2$ and $\alpha$ are $O(1)$ geometrical prefactors.)
At $a=b^2/(27 c)$, or $\gamma=2.7$ for the generic lyotropic, 
there is a first order transition to the nematic phases and 
as $\gamma$ is increased further the nematic order parameter $q$ increases.  
The variation of $q$ with $\gamma$ can be calculated analytically.  Agreement
with simulation results is excellent as shown in Figure 1.

Imposing a shear on the system in the nematic phase will act to align 
the director field along the flow gradient.  Assuming a steady-state, 
homogeneous flow and a uniaxial nematic state, it follows 
from (\ref{Qevolution}) that the angle between the 
direction of flow and the director, $\theta$, is given by \cite{BE94}
\begin{equation}
\xi \cos 2 \theta = \frac{3q}{2+q}.
\label{eq80}
\end{equation} 
The simulations reproduce this relation well as shown in Figure 2 for 
different values of $q$ and $\xi$.

When there is no solution to equation (\ref{eq80}) the director
tumbles in the flow or may move out of the plane to form a log-rolling
state\cite{DG93,RT98}. Figure 3 gives an example of this type of behavior,
showing the director angle as a function of time. 

Olmsted and Goldbart\cite{OG92} have argued that 
shear stress acts to favour the
nematic over the isotropic phase. Hence application of shear moves the phase
boundary, which extends from the first-order equilibrium transition at
zero shear along a line of first-order transitions which end at a
non-equilibrium critical point. Numerical results for this boundary are
shown in Figure 4.  The results are qualitatively similar to those of
\cite{OG92,OD97} who obtained the phase boundary for a
slightly different model using an interface stability argument.

On the coexistence line the liquid crystal prefers
to phase separate into shear bands
\cite{OG92,OD97,MR97}, 
coexisting regions of different strain rate running parallel to the
shear direction. Such shear banding occurs spontaneously in the
simulations reported here. An example is shown in Figure 5.

\section{Summary and Discussion}

In this paper we described in detail a lattice Boltzmann
algorithm to simulate liquid crystal hydrodynamics.  In the 
continuum limit we recover the Beris-Edwards formulation within which the 
liquid crystal equations of motion are written in terms of a
tensor order parameter. The equations are 
applicable to the isotropic, uniaxial nematic, and biaxial 
nematic phases.  Working within the framework of a variable tensor 
order parameter it is possible to simulate the dynamics of 
topological defects and non-equilibrium phase transitions between 
different flow regimes.

Lattice Boltzmann simulations have worked well for complex fluids 
where a free energy can be used to define thermodynamic equilibrium.  
However previous work has concentrated on self-assembly with much 
less attention being paid to more complex flow properties.  The 
algorithm described here includes
coupling between the order parameter and the flow. This allows the
investigation of non-Newtonian effects such as shear-thinning and 
shear-banding.  Examples are given in Section IV.

There are many directions for further research opened up by the rich
physics inherent in liquid crystal hydrodynamics and the generality of
the Beris-Edwards equations. For example results for liquid crystals 
under Poiseuille flow show that the director
configuration can depend on the sample history as well as the 
viscous coefficients and thermodynamic parameters\cite{DO00}.  The effect of 
hydrodynamics on phase ordering is being investigated\cite{DO00a} and it would be 
interesting to study the pathways by which different dynamic states 
transform into each other.  The addition of an electric field to the 
equations of motion will allow problems relevant to liquid crystal 
displays to be addressed. Numerical investigations are proving vital
as the complexity of the equations makes analytic progress difficult.

\section{Appendix A}

We outline how the Beris-Edwards equations reduce to those of
Ericksen, Leslie, and Parodi in the uniaxial nematic phase when the
magnitude of the order parameter remains constant. Hence we obtain
expressions for the Leslie coefficients in terms of the parameters
appearing in the equations of motion (\ref{Qevolution}) 
and (\ref{NS})\cite{BE94}.

Taking $\vec{n}$ to represent the order-parameter field the 
Ericksen-Leslie stress tensor and the equation of motion for the order
parameter are, respectively\cite{E66,L68,DG93},
\begin{eqnarray}
\sigma_{\alpha\beta}^{EL} &=& \alpha_1 n_\alpha n_\beta n_\mu n_\rho
D_{\mu\rho} + \alpha_4 D_{\alpha\beta}+\alpha_5 n_\beta n_\mu
D_{\mu\alpha} \nonumber \\
&+&\alpha_6 n_\alpha n_\mu D_{\mu\beta}+\alpha_2 n_\beta
N_\alpha+\alpha_3 n_\alpha N_\beta, 
\label{aaa}\\
h_\mu^{EL} &=& \gamma_1 N_\mu+\gamma_2 n_\alpha D_{\alpha\mu}
\label{heqn}
\end{eqnarray}
together with the relations
\begin{eqnarray}
\gamma_1&=&\alpha_3-\alpha_2, \\
\gamma_2&=&\alpha_6-\alpha_5=\alpha_2+\alpha_3.
\end{eqnarray}
The second of these, known as Parodi's relation, is a
result of Onsager reciprocity. 
(Note that, following the convention in (\ref{NS}), 
the stress tensor is written so that in the corresponding
Navier-Stokes equation one contracts on the second index when taking
the divergence.)

The $N_\alpha$ are co-rotational derivatives
\begin{eqnarray}
N_\alpha
&=& \partial_t n_\alpha+u_\beta \partial_\beta n_\alpha-\Omega_{\alpha\mu}n_\mu.
\end{eqnarray}
The molecular field ${\vec h}$ is given by
\begin{eqnarray}
h_\mu&=&-{\delta {\cal F} \over \delta n_\mu}\;\;=\;\; \kappa^{EL} \nabla^2 n_\mu + \zeta({\bf r}) n_\mu
\label{helastic}
\end{eqnarray}
where the last line assumes the one-elastic constant
approximation and $\zeta$ is a Lagrange multiplier to impose
${\vec n}^2=1$.

To obtain the Ericksen-Leslie-Parodi equations from the tensor formalism
uniaxial symmetry is imposed on the order parameter
\begin{equation}
Q_{\alpha\beta}=q (n_\alpha n_\beta -1/3 \delta_{\alpha\beta}).
\label{qdef}
\end{equation}
where $q$ is the magnitude of the largest eigenvalue.
We first obtain an expression for $\kappa^{EL}$ in terms of $\kappa$
and show that equation (\ref{H(Q)}) reduces to the form (\ref{helastic}).
Using the chain rule
\begin{eqnarray}
h_\mu^{EL}&=&-{\delta {\cal F} \over \delta n_\mu}\;\; = \;\;
 -{\delta {\cal F} \over \delta Q_{\alpha\beta}} {\partial
  Q_{\alpha\beta}\over \partial n_\mu}
\;\; = \;\;  q(H_{\mu\beta} n_{\beta}+ n_\alpha H_{\alpha\mu}).
\label{h(n,H)}
\end{eqnarray}
Substituting ${\bf H}$ from equation (\ref{H(Q)}) into equation (\ref{h(n,H)}), writing
${\bf Q}$ in uniaxial form and simplifying gives  after some
algebra 
\begin{equation}
h_\mu^{EL} = 2 q^2 \kappa \nabla^2 n_\mu.
\label{ELhelastic}
\end{equation}  
Terms proportional to $n_\mu$ have been omitted as these will only change
the magnitude of the order parameter and the Lagrange multipier $\zeta$
will adjust to prevent this. Hence comparing (\ref{helastic}) and (\ref{ELhelastic})
\begin{equation}
\kappa^{EL}= 2 q^2 \kappa.
\label{KEL}
\end{equation} 

Consider now the equation of motion for the order parameter (\ref{heqn}).
Solving the ${\bf Q}$-evolution
equation (\ref{Qevolution}) for ${\bf H}$, and writing ${\bf Q}$ in uniaxial
form gives
\begin{eqnarray}
\Gamma H_{\alpha\beta}&=& 
q (n_\beta
  N_\alpha+n_\alpha N_\beta)-q \xi(D_{\alpha\gamma}n_\gamma
  n_\beta+n_\alpha n_\gamma D_{\gamma\beta})\nonumber \\
& & \quad  +{2 \over 3}(q-1)\xi
  D_{\alpha\beta}+2 q^2 \xi n_\alpha n_\beta
  D_{\gamma\nu} n_\nu n_\gamma+{2 \over 3} q(1-q)\xi
  \delta_{\alpha\beta} D_{\gamma\nu} n_\nu n_\gamma.
\label{H(n,N,D)}
\end{eqnarray}
Substituting this into equation (\ref{h(n,H)}) yields,
after some algebra,
\begin{equation}
h_\mu=2 q^2 N_\mu-{2\over 3}q(q+2)\xi
n_\alpha D_{\alpha\mu}
\end{equation}
where we have again omitted terms proportional to $n_\mu$.  Comparison to
equation (\ref{heqn}) gives
\begin{eqnarray}
\gamma_1 &=& 2 q^2/ \Gamma, \\
\gamma_2 &=& -{2\over 3}q(q+2)\xi /\Gamma .
\label{gammas}
\end{eqnarray} 

Finally we consider how the stress tensor maps between the two
theories. Using equations (\ref{H(n,N,D)}) and (\ref{qdef}) the symmetric
(\ref{BEstress}) and antisymmetric (\ref{as}) 
parts of the Beris-Edwards stress tensor become, respectively,
\begin{eqnarray}
\Gamma \tau_{\alpha\beta}&=&q^2 (n_\alpha
  N_\beta-N_\alpha n_\beta)-q(q+2)/3 \xi (n_\alpha n_\gamma
  D_{\gamma\beta}-D_{\alpha\gamma} n_\gamma n_\beta)
\label{tauEL}\\
\Gamma \sigma_{\alpha\beta}&=&-\frac{q \xi}{3}(q+2)(n_\beta
  N_\alpha+n_\alpha N_\beta)+\frac{q \xi^2}{3}(4-q) (D_{\alpha\gamma} n_\gamma
  n_\beta+n_\alpha n_\gamma D_{\gamma\beta})
\nonumber \\
&&+ \frac{2 \xi^2}{3}(q-1)^2
  D_{\alpha\beta} 
  -\frac{8 q^2 \xi^2}{3}(\frac{3}{4}+q-q^2)\xi n_\alpha n_\beta
  D_{\gamma\nu} n_\nu n_\gamma 
\nonumber \\
&&
+\mbox{terms in}\;\;\;\;\;
\delta_{\alpha\beta} D_{\gamma\nu} n_\nu n_\gamma
\label{sigmaEL}
\end{eqnarray}
where we have ignored the final, distortion, term in (\ref{BEstress}).
A comparison of (\ref{tauEL}) and (\ref{sigmaEL}) to (\ref{aaa}) 
gives the Leslie coefficients (\ref{EL1})--(\ref{EL6}).
(These agree with the expressions given by Beris and Edwards in
\cite{BE94}, apart for the formula for $\alpha_1$. However the 
formula for $\alpha_1$ listed in \cite{BE90} is the same as that
calculated here.)

\section{Appendix B}

We obtain the second order term (\ref{eq27}) in the 
Chapman-Enskog expansion for the equation of motion of the order
parameter. Proceeding as in the derivation of (\ref{eq26}) but including
the second order term (\ref{eq25}) gives
\begin{equation}
\partial_t {\bf Q} + \partial_{\alpha} ({\bf Q} u_{\alpha}) - \widehat{\bf
H} =\tau_g \left\{ \partial_t^2 {\bf Q} +2 \partial_{\alpha} \partial_{t} ({\bf
Q} u_{\alpha}) + \partial_{\alpha} \partial_{\beta} ({\bf Q} u_{\alpha}
u_{\beta}) -\partial_t \widehat{\bf H} -\partial_{\alpha} (\widehat{\bf H} 
u_{\alpha}) \right\} 
\label{eqB1}
\end{equation}
where we have used the definitions (\ref{eq8}) and (\ref{eq9}) to perform
the sums over $i$. Equation (\ref{eq26}) shows that the first, half the
second and the fourth term in the curly brackets are together of
higher order in derivatives and can be eliminated.

We next note that 
\begin{equation}
\partial_{\alpha} \partial_{t}({\bf Q}u_{\alpha}) = 
\partial_{\alpha}\left ( -\frac{{\bf Q}}{\rho}(\partial_t  \rho) u_{\alpha} +(\partial_t 
{\bf Q}) u_{\alpha}+\frac{{\bf
Q}}{\rho}\partial_t  (\rho u_{\alpha})\right).
\label{intermediate}
\end{equation}
The time derivatives can be replaced by spacial derivatives by using
equations (\ref{eq36}), (\ref{eq26}), and (\ref{eq35})
respectively. Substituting back into equation (\ref{eqB1}) and
ignoring terms in $\sum_i p_i e_{i \alpha} \sim \partial_\beta
\tau_{\alpha\beta}$ that contain an extra derivative
\begin{eqnarray}
& &\partial_{t} {\bf Q} + \partial_{\alpha} ({\bf Q}u_{\alpha}) -
\widehat{\bf H} = 
\tau_g  \left\{ \partial_{\alpha} \left(
\frac{{\bf Q}}{\rho} \right) \partial_{\beta}(\rho u_{\beta})
u_{\alpha} 
- \partial_{\alpha}
\partial_{\beta} ({\bf Q} u_{\beta}) u_{\alpha}
+ \partial_{\alpha} (\widehat{\bf H} u_{\alpha}) \right\}
\nonumber\\ 
& & \qquad\qquad - \tau_g \left \{ \partial_{\alpha} \left(
\frac{{\bf Q}}{\rho} (\partial_{\beta} (\rho u_{\alpha} u_{\beta})
-\partial_{\beta} \sigma_{\alpha \beta}
) \right) +
\partial_{\alpha}\partial_{\beta}({\bf Q} u_{\alpha} u_{\beta}
) -  \partial_{\alpha}(\widehat{\bf H} u_{\alpha} ) \right\}.
\end{eqnarray}

Rearranging the derivatives this simplifies to
\begin{eqnarray}
\partial_{t} {\bf Q} + \partial_{\alpha} ({\bf Q}u_{\alpha}) -
\widehat{{\bf H}} &=& 
-\tau_g \left\{ \partial_{\alpha} \left( \frac {{\bf Q}}{\rho}
\partial_{\alpha} P_{0} \right) \right\}.
\end{eqnarray}

\begin{figure}
\centerline{\epsfxsize=3.in \epsffile{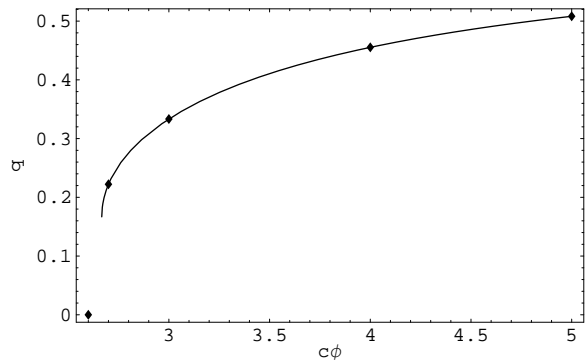}}
\vskip 0.2true cm
\caption{Equilibrium order parameter $q$ versus $c\phi$. The points
  are from a simulation and the line is the analytic result.}
\label{qeqfig}
\end{figure}

\begin{figure}
\centerline{\epsfxsize=3.in \epsffile{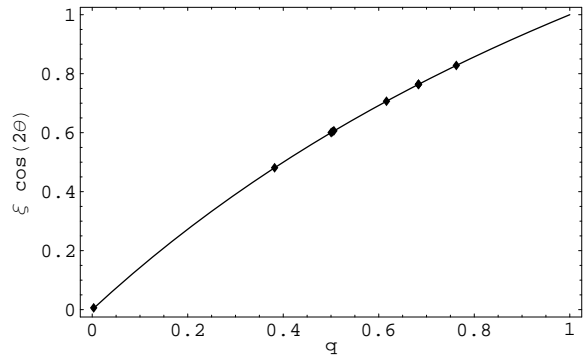}}
\vskip 0.2true cm
\caption{$\xi$ times the cosine of twice the angle between the
  director and the flow $\xi \cos(2 \theta)$ versus the magnitude of
  the order parameter $q$.  The points are from simulations and the
  line is the expected value $3 q/(2+q)$ from Equation (\ref{eq80}).}
\label{cos2tvsq}
\end{figure}

\begin{figure}
\centerline{\epsfxsize=3.in \epsffile{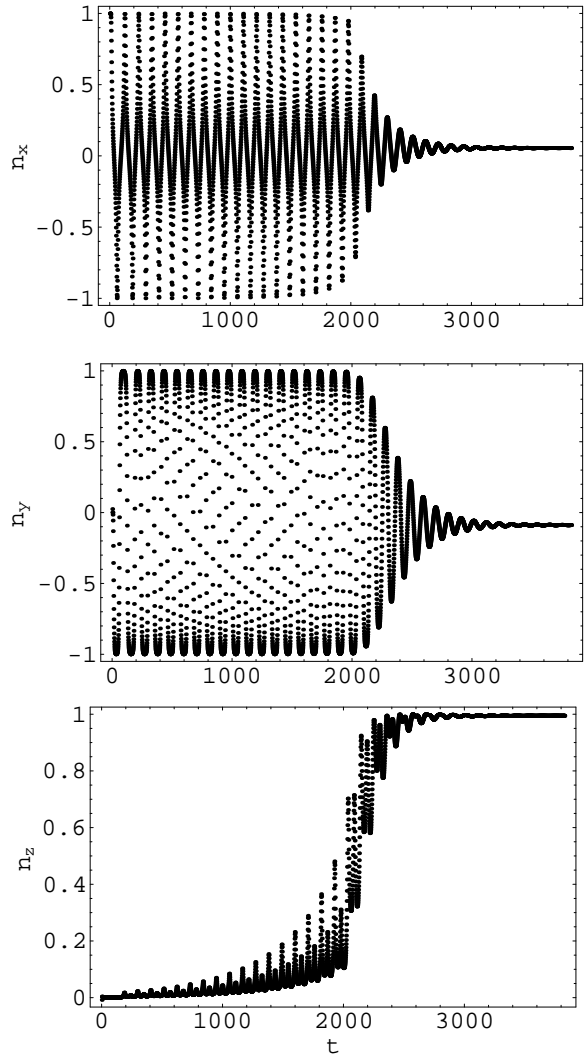}}
\vskip 0.2true cm
\caption{The components of the director as a function of time for a
  system changing from a metastable tumbling state to a stable
  log-rolling state.}
\label{log}
\end{figure}

\begin{figure}
\centerline{\epsfxsize=3.in \epsffile{fig4.epsi}}
\vskip 0.2true cm
\caption{Phase diagram in the shear stress $\Pi_{xy}$, effective
  temperature $a$ plane. ($a$ is the coefficient of the 
quadratic term in the free energy (\ref{free}).)}
\label{phasedia}
\end{figure}

\begin{figure}
\centerline{\epsfxsize=3.in \epsffile{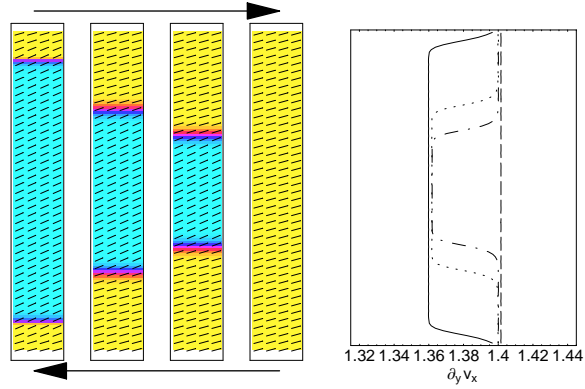}}
\vskip 0.2true cm
\caption{Shear bands for a range of strain rates.  The bands
are formed by the  coexistence of isotropic (darker) and
nematic states.  The variation of the strain rate across the system,
scaled by$100\Gamma$ to make it dimensionless, is also shown.}
\label{shrbnd}
\end{figure}

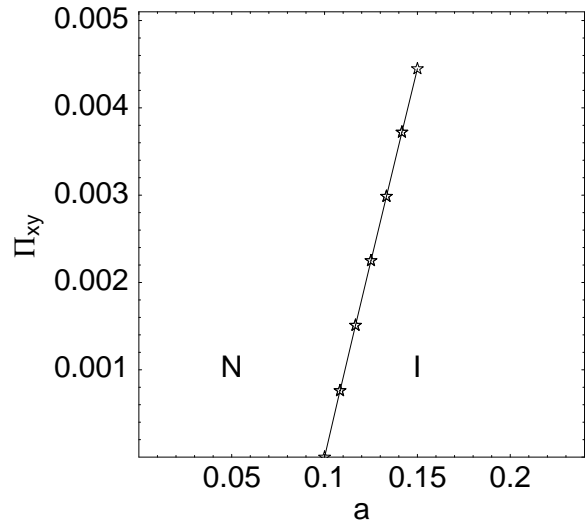
\end{document}